\documentclass[twocolumn,floatfix,prb,showpacs,superscriptaddress]{revtex4-1}

\usepackage{graphicx}
\usepackage{amssymb}
\usepackage{amsmath}
\usepackage{color}
\usepackage{psfrag}
\usepackage{epsfig}
\usepackage{bbm}
\usepackage{bm}
\usepackage{hyperref}
\hypersetup{breaklinks}

\newcommand{\ket}[1]{| #1 \rangle}

\newcommand{\rb}[1]{\left( #1 \right)}

\newcommand{\ew}[1]{\langle #1 \rangle}
\newcommand{\beq}{\begin{eqnarray}}
\newcommand{\eeq}{\end{eqnarray}}

\newcommand{\op}[2]{| #1 \rangle \langle #2 |}

\newcommand{\eq}[1]{Eq.~(\ref{#1})}
\newcommand{\fig}[1]{Fig.~\ref{#1}}

\newcommand{\trace}[1]{\mathrm{Tr}\left\{#1\right\}}

\begin{document}
\title{The Leggett-Garg inequality in electron interferometers}
\author{Clive Emary}
\affiliation{ Institut f\"ur Theoretische Physik,
  Technische Universit\"at Berlin,
  D-10623 Berlin,
  Germany}
\author{Neill Lambert}
\affiliation{Advanced Science
        Institute,
     The Institute of Physical and Chemical Research (RIKEN), Saitama 351-0198, Japan}
\author{Franco Nori}
\affiliation{Advanced Science
        Institute,
     The Institute of Physical and Chemical Research (RIKEN), Saitama 351-0198, Japan}
\affiliation{Physics Department, University of Michigan, Ann
Arbor, Michigan, 48109, USA}

\date{\today}
\begin{abstract}
  We consider the violation of the Leggett-Garg inequality in electronic Mach-Zehnder inteferometers. This set-up has two distinct advantages over earlier quantum-transport proposals: firstly, the required correlation functions can be obtained without time-resolved measurements.  Secondly, the geometry of an interferometer allows one to construct the correlation functions from ideal negative measurements, which addresses the non-invasiveness requirement of the Leggett-Garg inequality.  We discuss two concrete realisations of these ideas: the first in quantum Hall edge-channels, the second in a double quantum dot interferometer.
\end{abstract}
\pacs{
03.65.Ud,   %Bell inequalities,
73.23.-b, % Electronic transport in mesoscopic systems
03.65.Ta,   %Foundations of quantum mechanics; measurement theory
42.50.Lc % Quantum fluctuations, quantum noise, and quantum jumps
%73.63.Kv,  % Quantum dots (electronic transport
%73.50.Td,  % Noise processes and phenomena
%73.23.Hk   % Coulomb blockade; single-electron tunnelling
}
\maketitle
%%%%%%%%%%%%%%%%%%%%%%%%%%%%%%%%%%%%%%%%%%%%%%%%%%%%%%%%%%%%%%%%%%%%%%%%%

%%%%%%%%%%%%%%%%%%%%%%%%%%%%%%%%%%%%%%%%%%%%%%%%%%%%%%%%%%%%%%%%%
%%%%%%%%%%%%%%%%%%%%%%%   Introduction      %%%%%%%%%%%%%%%%%%%%%
%%%%%%%%%%%%%%%%%%%%%%%%%%%%%%%%%%%%%%%%%%%%%%%%%%%%%%%%%%%%%%%%%

Bell inequalities
set bounds on the nature of the correlations between
{\em spatially}-separated entities within local hidden variable
theories \cite{Bell1964,Bell2004}. In contrast, Leggett-Garg
inequalities (LGIs) set bounds on the {\em temporal} correlations of
a {\em single} system \cite{Leggett1985,Leggett2002}, and are
derived under the assumptions of {\em macroscopic realism} (MR)
and {\em non-invasive measurability} (NIM)
\footnote{
  From Ref.~\onlinecite{Leggett1985}, these assumptions are as follows. {\em Macroscopic realism:} A macroscopic system with two or more macroscopically distinct states available to it will, at all times, be in one or the other of these states. {\em Noninvasive measurability}: It is possible, in principle, to determine the state of the system with arbitrary small perturbation on its subsequent dynamics.  More information can be found in Refs.~\onlinecite{Leggett1985,Leggett2002}.
}
.

Bell and Leggett-Garg inequalities are related in that their
assumptions both imply the existence of a classical probability
distribution that determines experimental outcomes.
The probability amplitudes of quantum mechanics allow for
violation of these inequalities: with Bell, the violation is due to entanglement between
the two systems; with Leggett-Garg,
the violation occurs due to the superposition of system states and
their collapse under measurement.

The simplest LGI, henceforth referred to as {\em the} LGI, reads
\beq
  K \equiv C_{21} + C_{32} - C_{31} \le 1
  \label{K3intro}
  ,
\eeq where $C_{\alpha\beta} = \ew{Q(t_\alpha) Q(t_\beta)}$ is the
correlation function of the dichotomous variable $Q=\pm 1$ at
times $t_\alpha$ and $t_\beta$. Since the first experimental
violation \cite{Palacios-Laloy2010} of this inequality with weak measurements of a
superconducting qubit,  the Leggett-Garg
inequality has been experimentally probed in systems as diverse as
photons \cite{Goggin2011,Xu2011,Dressel2011}, defects in diamonds
centers \cite{Waldherr2011}, nuclear magnetic resonance
\cite{Athalye2011}, and phosphorus impurities in silicon
\cite{Knee2012}.
Whilst the subjects of these studies may not be macroscopic, the LGI performs a useful role for microscopic systems as an indicator that the device is operating beyond classical probability laws.
Moreover, if one accepts that the alternative to classical probabilities is quantum mechanics, the LGI provides a decisive indicator of the ``quantumness'' on a system \cite{Miranowicz2010}.

In this paper, we are interested in the violation of the LGI in quantum transport, and in particular, in electron-interferometers.
Although there has been much work on Bell inequalities in
electron transport, e.g. Refs~\onlinecite{Chtchelkatchev2002,Samuelsson2003,Beenakker2003,Samuelsson2004,Lebedev2005,Beenakker2006,Samuelsson2009,Emary2009,
Bednorz2011}, the LGI has only relatively recently been considered in
this setting \cite{Lambert2010, Emary2012}.  Specifically, the
charge flowing  through a confined nanostructure, e.g. double
quantum dot (DQD), has been shown to violate an inequality similar
to \eq{K3intro} out of equillibrium \cite{Lambert2010}.
Furthermore, the moment-generating function of charge transferred
through a device has also been shown to be subject to a set of
LG-style inequalities, which are violated for various quantum dot
models.  The violation of LGIs in  excitonic transport has also
attracted recent interest \cite{sun,Wilde2010}.

There are several difficulties which make the investigation of the LGI in electronic transport challenging in practice.
Ostensibly, the measurement of \eq{K3intro} requires time-resolved
measurements where the time between successive measurements is
smaller than the decoherence time of the system. For the
double quantum dot of Ref.~\onlinecite{Lambert2010}, for example, this
decoherence time is of the order of 1ns \cite{Hayashi2003}, which
makes the necessary time-resolved measurements very challenging (but, in principle, possible \cite{Petersson2010}).  Furthermore, for the
violation of \eq{K3intro} to be a meaningful indicator of
non-classical behaviour, it must be ensured that the measurements
are non-invasive.  This ``clumsiness loophole''\cite{Wilde2012}
that allows violations of \eq{K3intro} to be associated with
invasiveness of measurement, along with possible circumventions,
have been the subject of much discussion
\cite{Leggett1985,Ballentine1987,*Leggett1987,*Peres1988,*Leggett1989,*Tesche1990,*Elby1992,*Paz1993,*Benatti1995,*Calarco1995,*Onofrio1995,
Wilde2010}.

The transport set-ups we consider here are based on the electronic
Mach-Zehnder Interferometer (MZI), and can overcome both of these
problems.
The basic idea is that an electron travelling through a MZI can
take one of two paths, and this path index defines the variable
$Q=\pm 1$.  Unidirectional passage of the electron through the
system allows us to map the time indices of \eq{K3intro} onto
positions within the interferometer.  As we show below, this removes
the need for time-resolved measurements.

We consider two realisations of the MZI in which measurements of
$Q$ are performed in two different ways. In the first, the MZI is formed from quantum Hall edge channels, a set-up
which has been realised experimentally
\cite{Ji2003,Neder2006,Litvin2007,Roulleau2007,Neder2008,Roulleau2008,Roulleau2008a,Roulleau2009} and also attracted a
large degree of theoretical attention\cite{Seelig2001,Marquardt2004,Marquardt2004a,Foerster2005,Chung2005,Sukhorukov2007a,Foerster2007a,Foerster2007,Neder2007b,Neder2007a,Neder2007,Youn2008,Chang2008}
.  By interrupting the edge channels at various points and
diverting electron flow to current meters, we show that $K$ can
be obtained from   measurements of {\em mean currents} alone.
Furthermore, due to the spatial separation of the $Q=+1$ and
$Q=-1$ channels, our detectors interact with only one of the two
$Q$-states  at any given time.  Thus, our scheme provides a
natural way to implement {\em ideal negative measurements}, as
advanced by Leggett and Garg as a way to satisfy the NIM criterion
\cite{Leggett1985}.

The second set-up we consider is a MZI with a quantum dot (QD) in each arm. This
geometry is similar to several experiments \cite{Holleitner2001,Sigrist2004,Ihn2007,Muhle2008,Hatano2011} that have investigated transport through Aharanov-Bohm rings with QDs in the arms.  The difference here being that the dots are fed by two
tunnel-coupled leads \cite{Yamamoto2012}, rather than just one.
The QDs are monitored by quantum point contacts, whose transmission is sensitive to
the charge state of the QD
\cite{Gurvitz1997,Buks1998,Korotkov1999,Gurvitz2003,Averin2005,Ashhab2009}.
In contrast with the first set-up, electrons are not diverted out
of the MZI at any point, and the influence of the detectors occurs
as a pure dephasing effect.
The three correlation function in \eq{K3intro} are obtained
through a combination of mean currents, both through the MZI and
the quantum point contacts, and zero-frequency noise measurements, which
cross-correlate current fluctuations in the MZI and quantum point contacts.
As in the previous scheme, we construct an ideal negative measurement scheme with this set-up.

Both of these techniques exploit a combination of superpositions of
paths through an interferometer combined with a gathering of
``which-way'' information to violate the LGI.  The first set-up is
a particularly simple realisation of the LGI, and is by no means
restricted to transport, but could be used e.g. with photons,
atoms or molecules.

The paper proceeds as follows. In Sec.~\ref{SEC:MZI} we describe
the basics of testing the LGI in a MZI.  Sec.~\ref{SEC:QHE}
describes how this may be translated into experiments with quantum Hall effect
edge-channels.
Finally, Sec.~\ref{SEC:DQD} considers the alternative DQD-QPC geometry studied here.

%%%%%%%%%%%%%%%%%%%%%%%%%%%%%%%%%%%%%%%%%%%%%%%%%%%%%%%%%%%%%%%%%
%%%%%%%%%%%%%%%%%%%%%%%         MZI         %%%%%%%%%%%%%%%%%%%%%
%%%%%%%%%%%%%%%%%%%%%%%%%%%%%%%%%%%%%%%%%%%%%%%%%%%%%%%%%%%%%%%%%
\section{Mach-Zehnder interferometer \label{SEC:MZI}}

%%%%%%%%%%%%%%%%%%%%%%%%%%%%%%%%%%%%%%%%%%%%%%%%%%%%%%%%%%%%%%%%%
\begin{figure}[tb]
  \begin{center}
\includegraphics[width=0.7\columnwidth,clip]{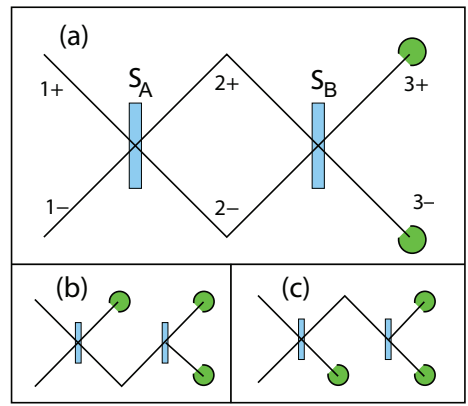}
  \caption{ (Color online)
  The Mach-Zehnder interferometer with three different detector configurations for the non-invasive measurement of the LGI of \eq{K3intro}.   Electrons are injected into the 1+ port.
  {\bf (a)} Complete MZI configuration with detectors only at the final outputs $3\pm$. With this set-up we can measure the probabilities $P^D_{3\pm}(1)$ and construct $C_{31}$.
  {\bf (b)} An additional detector is inserted into the MZI + arm.  With this configuration we can measure probabilities $P^D_{2+}(0)$ and $P^D_{3\pm;2+}(1,\cdot)$.
  {\bf (c)} A detector in the `$-$' arm  allows us to obtain $P^D_{2-}(0)$ and $P^D_{3\pm;2-}(1,\cdot)$.  Combining the results of (b) and (c) allows us to construct correlation functions $C_{21}$ and $C_{32}$.
    \label{MZIfig}
 }
  \end{center}
\end{figure}
%%%%%%%%%%%%%%%%%%%%%%%%%%%%%%%%%%%%%%%%%%%%%%%%%%%%%%%%%%%%%%%%%

We begin by describing an abstract version of our MZI scheme to outline the basic ideas.
The MZI is a two-channel interferometer with two beam-splitters that divide the MZI into three zones which we label: 1, the input ports; 2, the arms of the interferometer; and 3, the output ports (see \fig{MZIfig}).
We inject one electron at a time into the MZI and the path taken by the electron will be the degree-of-freedom under test with $Q=+1$ when the electron is located in the upper channel of the MZI, and $Q=-1$ the lower.
Since the electron passes sequentially through the three zones, we can map a measurement of $Q$ at time $t_\alpha$ to a ``which-way'' measurement at any point in the region $\alpha$ of the interferometer.
In particular,  $Q_1$ and $Q_3$ are measured at the input and output ports, and $Q_2$ is measured by placing detectors in the arms of the interferometer at $2\pm$, where $2$ refers to the zone, and $\pm$ the upper or lower channel.
In this section we assume that we have ideal single-electron detectors
that ``click'' on detecting an electron, which is then removed from the system (i.e., the detectors act essentially as electronic analogues of photodetectors).  More realistic measurements in terms of currents are discussed in section \ref{SEC:QHE}.

%%%%%%%%%%%%%%%%%%%%%
\subsection{Ideal negative measurements}
A detector placed in one of the arms interacts strongly with electrons in that path (they are completely removed from the MZI) and has no effect on electrons in the other.  With a detector placed at $Q=+1$, say, then the absence of a detector response (combined with MR and ideal detectors) allows us to infer the state of the system ($Q=-1$) without any disruption.
This is exactly the form of detector required to perform an ideal negative measurement as envisioned in Ref.~\onlinecite{Leggett1985}.

To make the measurement scheme as simple as possible, let us inject electrons into the $1+$ port, such that the initial state is known
\footnote{It is known that the initial state of the system does not affect the degree of violation of the LGI \cite{Knee2012a}}.  We do not need to measure in zone 1 and there is no question about the NIM of $Q_1$. The correlation function $C_{21}$ and $C_{31}$ boil down to measuring $\ew{Q_2}$ and $\ew{Q_3}$ respectively.

Let us define $P^D_{\alpha\pm}(n)$ as the probability that the
detector placed at position $\alpha\pm$ either detects ($n=1$) or
fails to detect ($n=0$) the electron. Since no further
measurements are made past point 3, it is irrelevant whether we
measure non-invasively or not at point 3.   Placing detectors at
$3\pm$, we measure the probabilities $P^D_{3\pm}(1)$, and the
$C_{31}$ correlation function can simplify be expressed as \beq
   C_{31} &=& P^D_{3+}(1) -  P^D_{3-}(1)
   \label{C31}
   .
\eeq
The set-up for this measurement is shown in \fig{MZIfig}a.

Since, in measuring $C_{21}$, no further measurements are made
after region 2, it is also not necessary to measure $C_{21}$
non-invasively.  We can measure $\ew{Q_2}$ (and thus $C_{21}$) by
running the experiment once with a detector in channel 2+, and
once in 2$-$ (\fig{MZIfig}b and c) and writing 
\beq
  C_{21} =  P^D_{2+}(1) -  P^D_{2-}(1). 
   \label{C21invasive}
\eeq
It is perhaps instructive to discuss how to obtain this
quantity using the ideal negative measurement technique and
measure $C_{21}$ in terms of the probabilities of absence of
detector clicks.
With the detector at $2+$, we can equate the probability that no
electron is detected, $P^D_{2+}(0)$, with the probability that the
electron travels the path $2-$.  Swapping the detector to the other
arm, we measure $P^D_{2-}(0)$ and infer the probability that the
electron takes path $2+$.  Whence, we obtain the non-invasively
measured 
\beq
   C_{21} &=&  P^D_{2-}(0) -  P^D_{2+}(0)
   \label{C21}
   .
\eeq
Since $P^D_{2\pm}(0) = 1-P^D_{2\pm}(1)$, \eq{C21invasive} and \eq{C21} give the same result.

We now consider $C_{32}$, where it is essential that we measure
$Q_2$ non-invasively, since a subsequent measurement is performed.
On the face of it, measuring $C_{32}$ requires a correlation
measurement between two detectors.  This, however, is not the
case, as we now show.

Let us begin by placing one detector at 2+ and another one at 3+
(\fig{MZIfig}b). We can then obtain the four probabilities,
$P^D_{3+,2+}(n,n')$, that the detectors at $3+$ and $2+$ give the
results $n,n'=0,1$  respectively. Of these, the one we are
interested in is  $P^D_{3+,2+}(1,0)$, since this allows us to infer
(non-invasively) the probability that the electron took path $2-$
to detector $3+$.  Moreover, we do not actually need to actively
detect at $2+$, since, if the electron reaches the 3+ detector, it
is clear that it has not entered channel 2+ (because the detector there would have 
removed the electron from the system)
\footnote{
  We do not actually need it to be a detector at 2$\pm$, anything that removes the electron from the system would suffice.  However, seeing as we need detectors for measuring $C_{21}$, it seems sensible to use these.
}.
With all four probabilities, $ P^D_{3q,2q'}(1,\cdot)$, obtained in this non-invasive way, we can construct
\beq
   C_{32} &=& - \sum_{q,q'=\pm} q q' P^D_{3q,2q'}(1,\cdot)
   \label{C32part}
   ,
\eeq
where we have replaced the measurement value at position 2 with a dot to indicate that we do not actually have to measure there (the value is guaranteed to be zero).

In this way we obtain all the required correlation functions,
measured non-invasively where necessary.  Although we have
concentrated on the simplest case here, the above non-invasive
techniques are extensible to the case where the input state is
unknown and all $C_{\alpha \beta}$ must be measured in a non-invasive way,
or to more complicated LGIs \cite{Athalye2011}.

%%%%%%%%%%%%%%%%%%%%%
\subsection{Leggett-Garg Inequality}

The action of the MZI can be specified by two beamsplitter
scattering matrices $s_X$;~$X=A,B$.  With $a_{\alpha q}$ the
annihilation operator for an electron in channel $\alpha q$, the
beamsplitter input-output relations read \beq
  \rb{
  \begin{array}{c}
   a_{2+} \\ a_{2-}
  \end{array}
  }
  =
  s_A
  \rb{
  \begin{array}{c}
   a_{1+} \\ a_{1-}
  \end{array}
  }
  ;\quad
  %\nonumber\\
    \rb{
  \begin{array}{c}
   a_{3+} \\ a_{3-}
  \end{array}
  }
  %&=&
  =
  s_B
  \rb{
  \begin{array}{c}
   a_{2+} \\ a_{2-}
  \end{array}
  }
  .
\eeq
Parameterizing the scattering matrices as
\beq
  s_X =
  \rb{
  \begin{array}{cc}
    \cos(\frac{1}{2}\theta_X)
      & \sin(\frac{1}{2}\theta_X)e^{i\frac{1}{2}\phi_X} \\
    -\sin(\frac{1}{2}\theta_X)e^{-i\frac{1}{2}\phi_X}
      & \cos(\frac{1}{2}\theta_X)
  \end{array}
  }
  ,
  %;
  %\quad
  %X = A,B
  \label{Sparam}
\eeq
we obtain the correlation functions
\beq
  C_{21} &=& \cos \theta_A
  ;
  \label{C21angles}
  \\
  C_{31} &=&\cos \theta_A \cos \theta_B- \sin \theta_A \sin \theta_B \cos\phi
  ;
  \label{C31angles}
  \\
  C_{32} &=&\cos \theta_B
  \label{C32angles}
  ,
\eeq
such that the LG correlator reads
\beq
  K(\theta_A,\theta_B,\phi) &=& \cos \theta_A + \cos \theta_B - \cos \theta_A \cos \theta_B
  \nonumber\\
  &&
  + \sin \theta_A \sin \theta_B \cos\phi
  \label{K3ideal}
  ,
\eeq
with $\phi = \frac{1}{2}(\phi_A-\phi_B)$ being the phase difference accumulated between the two paths.
This is a familiar expression.  If we identify $\theta_A = \Omega
\tau_1$ and  $\theta_B = \Omega \tau_2$, then  \eq{K3ideal} is exactly that obtained for a qubit evolving 
under the Hamiltonian $H=\frac{1}{2}\Omega \sigma_x$ measured in
the $\sigma_z$ basis at times $t_1$, $t_2=t_1+\tau_1$, and
$t_3=t_2+\tau_2$.  The properties of \eq{K3ideal} are discussed in
Sec.~\ref{SEC:QHE}

%%%%%%%%%%%%%%%%%%%%%%%%%%%%%%%%%%%%%%%%%%%%%%%%%%%%%%%%%%%%%%%%%
%%%%%%%%%%%%%%%%%%%%%         QHE            %%%%%%%%%%%%%%%%%%%
%%%%%%%%%%%%%%%%%%%%%%%%%%%%%%%%%%%%%%%%%%%%%%%%%%%%%%%%%%%%%%%%%
\section{Quantum Hall edge-channels \label{SEC:QHE}}

\begin{figure}[tb]
  \includegraphics[width=1.0\columnwidth,clip]{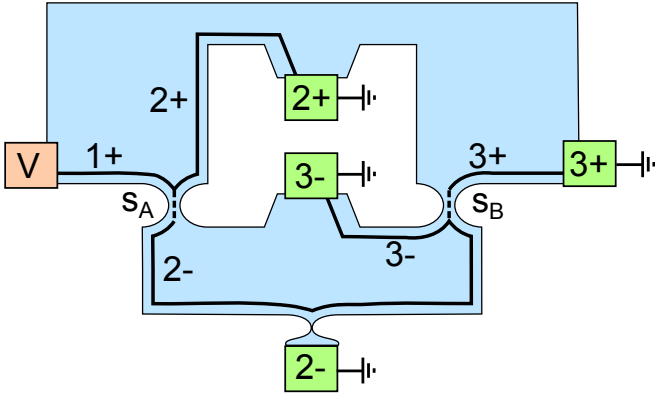}
  \caption{ (Color online)
  Quantum Hall edge-channel realisation of the MZI set-up for measurement of Legget-Garg inequality.  The MZI set-up is similar to that of Ref.~\onlinecite{Ji2003} but with two extra detectors ($2\pm$).
   These additional detectors can be isolated from the circuit by closing off the QPCs between them and the edge channel.
  The configuration shown has the detector at the 2+ position active such that transmission to beamsplitter B via channel 2+ is blocked, and the detector at $2-$ is pinched off.  This detector combination corresponds to that of \fig{MZIfig}b.
  \label{QHEfig}
 }
\end{figure}

Quantum Hall edge channels have been shown to allow a direct
translation of the MZI into electronic transport experiments
\cite{Ji2003,Neder2006,Litvin2007,Roulleau2007,Neder2008,Roulleau2008,Roulleau2008a,Roulleau2009} and \fig{QHEfig}
shows a sketch of the quantum Hall geometry needed to realise our
proposal. Each channel in the MZI is realised with a single
edge-channel and the electronic beam-splitters are realised by
quantum point contacts (QPCs).  Backscattering is  suppressed
between edge-channels such that transport is unidirectional.  This
set-up is the same as the MZIs of experiment except for the
addition of extra contacts to the arms of the interferometer.
These contacts are connected to the edge-channels via adjustable
quantum point contacts, such that the detectors can be coupled
into and out of the MZI as required.  This method of coupling probes to the MZI arms has been realised in Ref.~\onlinecite{Roulleau2009}.
Port 1+ is raised to a voltage $+V$ and electrons are injected into this channel.
The output ports (detectors) are all grounded.  When the correlation function $C_{31}$ is being measured, the detectors at $2\pm$ are not required and are isolated from the MZI by closing their QPCs.
(\fig{QHEfig} shows detector $2-$ closed off in this way).  To
measure the remaining correlation functions, the detectors at
$2\pm$ are, one then the other, connected into the MZI by opening
up their respective QPCs.  In \fig{QHEfig}, the detector at $2+$
is connected into the circuit and fully prevents electrons in
channel $2+$ from reaching the outputs $3\pm$.

%%%%%%%%%%%%%%%%%%%%%%%%%%%%%%%%%%%%%%%%%%%%%%%%%%%%%%%%%%%%%%%%%
%%%%%%%%%%%%%%%%%%%%%%%%%%%%%%%%%%%%%%%%%%%%%%%%%%%%%%%%%%%%%%%%%

%%%%%%%%%%%%%%%%%%%%%%%%%%%%%%%%%%%%%%%%%%%%%%%%%%%%%%%%%%%%%%%%%
%%%%%%%%%%%%%%%%%     Current measurements      %%%%%%%%%%%%%%%%%
%%%%%%%%%%%%%%%%%%%%%%%%%%%%%%%%%%%%%%%%%%%%%%%%%%%%%%%%%%%%%%%%%
\subsection{Current measurements}

\begin{figure}[tb]
  \begin{center}
    \includegraphics[width=\columnwidth,clip]{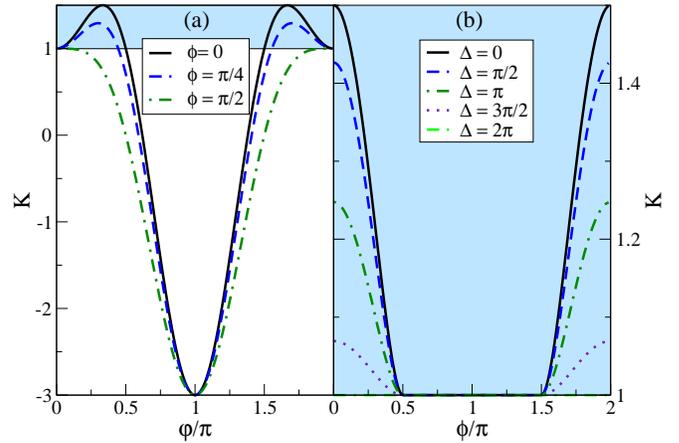}
    \caption{ (Color online)
      {\bf (a)}
      LG correlator $K(\theta_A,\theta_B,\phi)$ of \eq{K3ideal} as a function of
       the beamsplitter angle $\theta=\theta_A=\theta_B$ for three values of the phase $\phi=0,\pi/4,\pi/2$.  The shaded blue region indicated violation of the LG inequality ($K>1$).  The maximum violation, $K_\mathrm{max} = \frac{3}{2}$, occurs for $\phi=0$ and, e.g., $\theta=\pi/3$.
      {\bf (b)}  The influence of dephasing.  Shown is the
      LG correlator $K$ of \eq{K3dephasing} maximized over $\theta_{A/B}$ as a function of the phase $\phi$.
      Results shown for values of the dephasing parameter $\Delta/\pi = 0,\frac{1}{2},1,\frac{3}{2},2$.
      Violations of the LG are only observed for $\cos\phi >0$.
    \label{MZIresults}
   }
  \end{center}
\end{figure}

Let $\ew{I_{\alpha q}}$ be the mean stationary current flowing
into output $\alpha q$, given that when $\alpha=3$, the detectors
at positions $2\pm$ are closed off.  Further, let
$\ew{I_{3q;2q'}}$ be the current flowing at output $3q$ when the
output at $2q'$ is open. Since, in the linear regime, the current
operator for each output is  $I_{\alpha q}=G_0 V a^\dag_{\alpha q}
a_{\alpha q}$, with $G_0=e^2/h$ the conduction
quantum\cite{Blanter2000}, these mean currents are proportional to
the probability that an electron travels in the corresponding
channel.
The correlation functions required for the LGI can then be
constructed, as with the CHSH inequality \cite{Clauser1969,Samuelsson2003,Beenakker2003}, as
\beq
  C_{\alpha 1} &=& \frac{\ew{I_{\alpha +}} -  \ew{I_{\alpha -}}  }{\ew{I_{\alpha +}} + \ew{I_{\alpha -}}}
  ;
  \label{Ci1I}\\
  C_{32} &=&
  \frac{
    -\sum qq'\ew{I_{3q;2q'}}
  }{
    \sum \ew{I_{3q;2q'}}
  }
  \label{C32I}
   .
\eeq Division by the sum of detector currents removes
proportionality factors and, if all detector are identical, also
removes detector inefficiencies. Writing the scattering matrices
as
\beq
  s_X
  =
  \rb{
  \begin{array}{cc}
    r_X & t'_X \\
    t_X & r'_X
  \end{array}
  }
  \label{scatrt}
  ,
\eeq
we obtain the correlation functions
\beq
  C_{21} &=& |r_A|^2 - |t_A|^2;
  \nonumber\\
  C_{31} &=& |r_B r_A + t'_B t_A|^2 - |t_B r_A + r'_B t_A|^2;
  \nonumber\\
  C_{32} &=&
  |r_A|^2 \left\{|r_B|^2 - |t_B|^2\right\}
  - |t_A|^2\left\{|t'_B|^2 - |r'_B|^2\right\}
  .~~~
  \label{Cfns_scat}
\eeq
With scattering matrices as in the previous section, the LG parameter $K$ obtained from current measurements is the same as \eq{K3ideal}.
This quantity is plotted in Fig.~\ref{MZIresults}a.  A maximum violation of $K_\mathrm{max}=\frac{3}{2}$ is obtained for parameters $\theta_A=\theta_B=\pi/3$ and $\phi=0$.

The violation the LGI in this set-up arises because the
measurements at 2$\pm$ remove electrons from the interferometer
arms, preventing interference between the two paths.  The presence
of this interference in $C_{31}$ combined with its absence in
$C_{32}$ leads to the violation.

%%%%%%%%%%%%%%%%%%%%%%%%%%%%%%%%%%%%%%%%%%%%%%%%%%%%%%%%%%%%%%%%%
%%%%%%%%%%%%%%%%%%%%%%%%%%%%%%%%%%%%%%%%%%%%%%%%%%%%%%%%%%%%%%%%%

%%%%%%%%%%%%%%%%%%%%%%%%%%%%%%%%%%%%%%%%%%%%%%%%%%%%%%%%%%%%%%%%%
%%%%%%%%%%%%%%%%%%%%%      dephasing         %%%%%%%%%%%%%%%%%%%
%%%%%%%%%%%%%%%%%%%%%%%%%%%%%%%%%%%%%%%%%%%%%%%%%%%%%%%%%%%%%%%%%
\subsection{Dephasing}

We can account for the effects of dephasing by allowing the phase $\phi$ to fluctuate.  We replace $\phi\to\phi+\delta\!\phi$ in \eq{K3ideal} and integrate $\delta\!\phi$ over a flat distribution
in the range $-\Delta/2<\delta\!\phi<\Delta/2$.  The resulting LG parameter with dephasing reads
\beq
  K^\mathrm{deph}
  &=&
  \cos \theta_A + \cos \theta_B - \cos \theta_A \cos \theta_B
  \nonumber\\
  &&
  + f(\Delta)\sin \theta_A \sin \theta_B \cos\phi
  \label{K3dephasing}
  ,
\eeq
with $f(\Delta) = 2\Delta^{-1}\sin(\Delta/2)$ the function containing the dephasing effects
\footnote{This function $f(\Delta)$ is the amplitude of AB oscillations (normalised to the maximum current possible) that would be observed by sweeping the magnetic flux through the device with $\theta_A=\theta_B=\pi/2$.}.
If all angles are freely variable then the maximum of this function is
\beq
  K^\mathrm{deph}_\mathrm{max}(\Delta)
  =
  \frac{1+  f(\Delta)(1+f(\Delta))}{1+f(\Delta)}
  ,
\eeq
obtained for $\cos\theta_A=\cos\theta_B = [1+f(\Delta)]^{-1}$.
Expanding for small $\Delta$, we find $K^\mathrm{deph}_\mathrm{max}(\Delta) =\frac{3}{2}-\frac{1}{32}\Delta^2 $.
In the opposite limit, where the dephasing is total, $\Delta\to2\pi$, we have $f(\Delta)\to0$ and the maximised Leggett-Garg correlator reverts to the classical value, $ \lim_{\Delta \to 2\pi}K^\mathrm{deph}_\mathrm{max}= 1$ as required.

One interesting feature occurs if we assume that the phase $\phi$ is fixed (e.g., we are not able to vary the magnetic field) and maximise over $\theta_{A/B}$ (see Fig.~\ref{MZIresults}b).
Providing that $\cos\phi>0$, the maximum value is
\beq
  K^\mathrm{deph}_{\mathrm{max}(\theta_{{A/B}})}(\phi,\Delta)
  =
  f(\Delta)\cos\phi + \frac{1}{1+f(\Delta)\cos\phi}
  ,
\eeq
found by setting  $\cos\theta_A=\cos\theta_B = [1+f(\Delta)\cos\phi]^{-1}$.  If, however, $\cos \phi\le 0$, the maximum value is just the classical value,
$
 K^\mathrm{deph}_{\mathrm{max}(\theta_{{A/B}})} =1
$, found by setting $\cos\theta_A=\cos\theta_B =1$. This reversion
to the classical value occurs when the scalar product between the axis of the rotation of beamsplitter $B$ and that of beamsplitter $A$ becomes negative.

%%%%%%%%%%%%%%%%%%%%%%%%%%%%%%%%%%%%%%%%%%%%%%%%%%%%%%%%%%%%%%%%%
%%%%%%%%%%%%%%%%%%%%%        Multichoice      %%%%%%%%%%%%%%%%%%%
%%%%%%%%%%%%%%%%%%%%%%%%%%%%%%%%%%%%%%%%%%%%%%%%%%%%%%%%%%%%%%%%%
\subsection{Multi-channel case \label{SEC:MULTI}}

The above scheme is easily modified to include multiple channels.
We take the same geometry as before but assume that each lead
supports $M$ channels.  The $M$ channels of the upper lead are
all associated with qubit state $Q=+1$; the $M$ channels in the
lower lead, with state $Q=-1$.  The scattering matrices of
\eq{scatrt} are thus generalised to $2M\times 2M$ matrices with
$M\times M$ blocks, $r_X$, $t_X$, $r'_X$, and $t'_X$.
Assuming a large source-drain voltage, such that all channels are
equally populated, the correlation functions read: \beq
  C_{21} &=& \textstyle{\frac{1}{M}}
  \trace{R_A - T_A}
  ;
  \nonumber\\
  C_{32} &=& \textstyle{\frac{1}{M}}
  \trace{
    R_A^\dag \rb{R_B - T_B }
    + T_A^\dag \rb{R_B' - T_B'}
  }
  ;
  \nonumber\\
  C_{31} &=& \textstyle{\frac{1}{M}}
  \trace{
    R_A^\dag \rb{R_B - T_B }
    - T_A^\dag \rb{R_B' - T_B'}
  }
  \nonumber\\
  &&
  +
  \textstyle{\frac{1}{M}}
  \trace{
    r_A t_A^\dag\rb{{t_B'}^{\!\!\dag} r_B - {r'_B}^{\!\!\dag} t_B}
  \right.
  \nonumber\\
  && ~~~~ ~~~~ ~~~~
  \left.
    + t_A r_A^\dag \rb{r_B^\dag t_B' - t_B^\dag r_B'}
  },
\eeq
with $R_A = r_A^\dag r_A$,  $T_A = t_A^\dag t_A$, etc.  The
second term in the expression for $C_{31}$ arises from
interference between the paths.  In the single channel case, these
results reduce to those of \eq{Cfns_scat}.

An important observation can be made about the multi-channel case
by considering that the scattering matrices preserve the
channel-index, i.~e. we essentially have $M$ independent
interferometers. In this case, the LG parameter reads $
  K = \textstyle{\frac{1}{M}}  \sum_{m=1}^M K^{(m)}
$, where $K^{(m)}$ is the LG parameter for channel $m$. If we
could tune by hand all the parameters of the scattering matrices,
then the maximum violation of $K_\mathrm{max}=\textstyle{\frac{3}{2}}$ can be
reached. However, in an experiment, there will typically only be a
few controllable parameters and this could make violations hard to
observe.  Let us assume that we can adjust the parameters such
that one of the $K^{(m)}$ is maximised, $m=1$, say.  Whether we
see a violation or not very much depends on what happens with the
parameters of the other channels. If these parameters are all
roughly similar to those of channel 1, then violations should
still be observed. Generically, however, this will not be the
case, and the $K^{(m)}$ for the other channels will take unrelated
values in the range from $-3$ to $\frac{3}{2}$.  The negative
values are particulary troublesome as they will tend to overwhelm any
positive contribution to the violation from other channels.  This
lack of controllability means that multi-channel geometries are
best avoided if violations of the LGI are sought.

%%%%%%%%%%%%%%%%%%%%%%%%%%%%%%%%%%%%%%%%%%%%%%%%%%%%%%%%%%%%%%%%%
%%%%%%%%%%%%%%%%%%%%%        DQD        %%%%%%%%%%%%%%%%%%%
%%%%%%%%%%%%%%%%%%%%%%%%%%%%%%%%%%%%%%%%%%%%%%%%%%%%%%%%%%%%%%%%%
\section{Double quantum dot interferometer \label{SEC:DQD}}

\begin{figure}[tb]
  \begin{center}
    \includegraphics[width=0.7\columnwidth,clip]{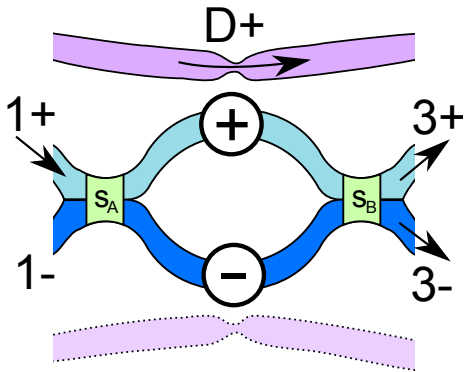}
  \caption{ (Color online)
         Sketch of a Mach-Zehnder interferometer with a quantum dot in each arm.
         The charge state of each QD can be monitored by the currents flowing through QPCs next to the dots. Here only the QPC monitoring the dot $+$ is active, such that the correlation function to be measured is as \fig{MZIfig}b.
    \label{DQDfig}
 }
  \end{center}
\end{figure}

The above MZI scheme functions by having the detectors remove
electrons from the interferometer arms.  In this section we study
a second MZI realisation which leaves the electrons within the
system and the effects of measurement are only felt through
dephasing.
This second set-up is shown in  \fig{DQDfig}.  As in the
foregoing, the basic structure is of two (single-channel) leads that are joined at two  beamsplitters.
Beamsplitters between non-edge-channel leads can be realised by
tunnel junctions, as in the recent experiments by Yamamoto
{\em et al.}\cite{Yamamoto2012}.
In each arm of the interferometer there is a QD and alongside each QD is
a QPC charge detector.  When connected to a voltage supply, the
current flowing through the QPCs serves as read-out of the
occupation of their respective QDs.  Note that although similar
detectors we used in e.g.
Refs.~\onlinecite{Ruskov2006,Jordan2006}, the way in which they
are used  here is different.

%%%%%%%%%%%%%%%%%%%%%%%%%%%%%%%%%%%%%%%%%%%%%%%%%%%%%%%%%%%%%%%%%
%%%%%%%%%%%%%%%%%%%%%%%%%%%%%%%%%%%%%%%%%%%%%%%%%%%%%%%%%%%%%%%%%

\subsection{Model}
We first consider the system without detectors.  Our MZI model is
related to that of, e.g., Refs.~\onlinecite{Marquardt2003,Urban2009}, but with different leads.
Far from the junctions, we describe the four leads as non-interacting Fermi reservoirs with Hamiltonian
$
  H_\mathrm{res} = \sum \omega_{k\alpha q}c^\dag_{k \alpha q}c_{k \alpha q}
$ with $k$ the wavenumber of the electron, and where $\alpha =
1,3$ and $q=\pm$ specify the lead (we set $\hbar=1$ and ignore
spin). We assume that there is but a single orbital of relevance
in each dot and that the DQD system is in the strong Coulomb
blockade regime, such that it is restricted to just three states:
`empty', $\ket{0}$; or with one excess electron in either the upper or
lower dots, $\ket{+} = d^\dag_+\ket{0}$ and $
\ket{-}=d^\dag_-\ket{0}$, respectively. Assuming the dot levels
are detuned by an energy $\epsilon$ from one another, the dot
Hamiltonian reads $
  H_S = \frac{1}{2}\epsilon \sum_q q d^\dag_q d_q
$.  In the following, we set $\epsilon=0$ for simplicity.

We assume that the effect of the beam-splitters is to modify the
amplitudes with which the leads couple to the QDs. So, for
example, an electron in lead 1+ tunnels into a superposition of
upper and lower dot states, with the details of the superposition
being determined by the scattering matrix $s_A$. The tunnel
Hamiltonian connecting lead and dots therefore reads \beq
  H_\mathrm{T}
  =
  \sum_k 
  \rb{
  \mathbf{C}^\dag_{k1} \cdot s_A^\dag \cdot \mathbf{d}
  +
  \mathbf{C}^\dag_{k3} \cdot s_B \cdot \mathbf{d}
  + \mathrm{H.c.}
  }
  ,
\eeq
where $s_{A,B}$ are scattering matrices,
assumed to be energy independent, $\mathbf{d} = (d_+,d_-)$ is a
vector of dot operators, and $ \mathbf{C}^\dag_{k\alpha} =
(T_{\alpha+}c^\dag_{k\alpha+},T_{\alpha-}c^\dag_{k\alpha-})$ are
vectors of lead operators with tunnel matrix elements $T_{\alpha
s}$, also assumed to be energy-independent. The corresponding
sequential tunnel rates are $\Gamma_{\alpha  q} = 2\pi |T_{\alpha
q}|^2 \varrho_{\alpha q}$, where $ \varrho_{\alpha q}$ is the
density-of-states of reservoir $\alpha q$, also assumed constant.

In the infinite-bias limit, the system can be described by a
quantum master equation of Lindblad-form
\cite{Gurvitz1998,Bagrets2003,Brandes2005}.  Let us introduce the
super-operator notations $
  \mathcal{J}[d]\rho = d\rho d^\dag
$
and
$
  \mathcal{A}[d]\rho = - \frac{1}{2} \left\{d^\dag d,\rho\right\}
$\cite{Gardiner1993,Carmichael1993a,Wiseman1994}, and introduce the operators
\beq
  \tilde{d}_{1q} = \sqrt{\Gamma_{1q}} \mathbf{e}_q \cdot s_A \cdot \mathbf{d}
  ;\quad
  %\nonumber\\
  \tilde{d}_{3q} = \sqrt{\Gamma_{3q}} \mathbf{e}_q \cdot s_B \cdot \mathbf{d}
  ,
\eeq
with unit vectors $\mathbf{e}_+=(1,0)$ and
$\mathbf{e}_-=(0,1)$. With introduction of counting fields
$\chi_{\alpha q}$ to facilitate the calculation of current
statistics (see e.g. Refs.~\onlinecite{Levitov1996,*Levitov2002,Bagrets2003}), the
$\chi$-resolved master equation for the DQD system reads
\beq
  \dot \rho(\chi) &=& -i \left[H_S,\rho\right]
  +
  \sum_{\alpha q}
  \rb{
    e^{i\chi_{\alpha q}}
    \mathcal{J}[\tilde{d}_{\alpha q}]
    -
    \mathcal{A}[\tilde{d}_{\alpha q} ]
  }\rho
  \label{QME}
  .~~~
\eeq

The QDs are monitored by QPCs in a set-up similar to the single
dot in an interferometer in the experiment of
Ref.~\onlinecite{Buks1998}. In including the detectors in our
theory, we follow Gurvitz \cite{Gurvitz1997,Gurvitz2003}.
When dot $q$ is unoccupied, the Hamiltonian for QPC $Dq$ reads
\beq
  H_{Dq}
  =
  \sum_{k s}\omega^D_{ks q} a^\dag_{ks q} a_{ks q}
  +\Omega_q \sum_k  a^\dag_{k L q} a_{k R q} + \mathrm{H.c}
  ,~~~
\eeq where $\omega^D_{ks q}$ is the energy of an electron in state
$k$ on side $s=L,R$ of the the QPC, and $\Omega_q$ is the coupling
amplitude between the two sides, (assumed energy independent).
When dot $q$ is occupied, we assume that this Hamiltonian is
modified such that the coupling constants shift to different
values, $\Omega_q \to \Omega_q'$.
In the limit of large bias across the QPC, the detector at
location $q$ gives rise to an extra Liouvillian
\beq
  \mathcal{W}_{Dq}(\chi_{Dq}) =
    e^{i\chi_{Dq}}
    \mathcal{J}[\tilde{d}_{Dq}]
    -
    \mathcal{A}[\tilde{d}_{Dq}]
    ,
\eeq
which adds to the DQD Liouvillian.  Here, $\tilde{d}_{Dq} = \sqrt{\gamma'_q}\op{q}{q} + \sqrt{\gamma_q}\rb{1-\op{q}{q}}$ with $\gamma_q$ the rate of electron transfer through the QPC $q$ when its dot is empty, and $\gamma_q'$ the rate when the dot is occupied. The counting field $\chi_{Dq}$ here allows us to calculate the statistics of the detector currents.
Microscopically, the rates are
$
  \gamma_q= 2\pi |\Omega_q|^2\rho_{Lq} \rho_{Rq} V_{Dq}
$
and
$
  \gamma_q'= 2\pi |\Omega'_q|^2\rho_{Lq} \rho_{Rq} V_{Dq}
$,
with $\rho_{s q}$ the density of states of the QPC reservoir $s q$ and $V_{Dq}$ the applied voltage.
Detectors may be decoupled or coupled from the MZI-QD system by
adjusting the QPC voltages such that the differences between the
amplitudes $\Omega_q$ and $\Omega_q'$ is either zero (decoupled)
or finite (coupled).  Here, we only couple at most one
detector to the system at a given time.  Furthermore, we assume
balanced detectors such that with the $D+$ detector coupled we
have $\gamma_+=\gamma$, $\gamma_+'=\gamma'$, and
$\gamma_-=\gamma_-'=0$, and when the $D-$ detector is coupled we have
$\gamma_-=\gamma$, $\gamma_-'=\gamma'$, and
$\gamma_+=\gamma_+'=0$,

\subsection{Current, correlation functions and probabilities}

Our approach to measuring the LGI with this set-up is similar to that with the quantum Hall edge-channels with the main exception being how $C_{32}$ is obtained.
We inject electrons into the `$+$'-channel of lead 1 and close the
`$1-$' channel: $\Gamma_{1+}\to\Gamma_L$ and  $\Gamma_{1-}\to0$.
For simplicity, we set the output rates equal: $\Gamma_{3+} =
\Gamma_{3-} = \Gamma_R$.

To obtain $C_{31}$, we switch off the QPC detectors and measure the output currents at $3\pm$.   Arranging the elements of the density matrix into a vector in the basis $(00,++,--,+-,-+)$, the stationary state of the DQD system reads
\beq
  \rho_\mathrm{stat} &=& \frac{1}{2 \Gamma}
  \rb{\frac{}{}
    2\Gamma_R,
    \Gamma_L(1+\cos\theta_A),
    \Gamma_L(1-\cos\theta_A),
  \right.
  \nonumber\\
  &&
  ~~~~
  \left.
    -e^{i\phi_A/2}\Gamma_L\sin\theta_A,
    -e^{-i\phi_A/2}\Gamma_L\sin\theta_A
  }
  ,
\eeq
with total width $\Gamma=\Gamma_L+\Gamma_R$.  Here, we have
assumed the same scattering matrices as in \eq{Sparam}. The
total current flowing is $\ew{I}_\mathrm{tot} = \ew{I}_{1+}
=\Gamma_L\Gamma_R/\Gamma$, which is divided between the output
ports as \beq
  \ew{I}_{3\pm} = \frac{\ew{I}_\mathrm{tot}}{2}
  \rb{
    1 \pm \cos \theta_A\cos \theta_B
    \mp \cos\phi\sin \theta_A\sin \theta_B
  }
  \nonumber
  .
\eeq
Constructing $C_{31}$ as in \eq{Ci1I} we obtain
\beq
  C_{31}
  =\cos \theta_A\cos \theta_B
    - \cos\phi\sin \theta_A\sin \theta_B
    ,
\eeq
which agrees with that of \eq{C31angles}.

Next we can obtain $C_{21}$ by turning on the QPC detectors one at
a time. As shown in Ref.~\onlinecite{Gurvitz1997}, the mean current flowing
through the QPC can be used to extract the mean current flowing
through the corresponding dot.  With a detector coupled to dot $q$,
the current through the detector is
\beq
  \ew{I_{Dq}}= \frac{\ew{I}_\mathrm{tot}}{2\Gamma_R}
  \left[
    \gamma\rb{1+2\frac{\Gamma_R}{\Gamma_L}}
    +\gamma'
    +q (\gamma'-\gamma) \cos\theta_A
  \right]
  \nonumber
  .
\eeq
The current flowing through the QPC when the DQD is empty is
$
  \ew{I_{Dq}^0}=\gamma
$,
such that the difference is
\beq
   \ew{\Delta I_{Dq}} &=& \ew{I_{Dq}} -  \ew{I_{Dq}^0}
   \nonumber\\
   &=&
   \frac{\gamma_q' - \gamma_q}{2\Gamma_R}\ew{I}_\mathrm{tot}
   \rb{1+q \cos \theta_A}
   ,
\eeq
which is proportional to the probability that an electron takes the path $2q$. Assuming balanced detectors, we obtain
\beq
  C_{21} =
  \frac{\ew{\Delta I_{D+}} - \ew{\Delta I_{D-}}}
    {\ew{\Delta I_{D+}} + \ew{\Delta I_{D-}}}
    =
    \cos \theta_A
    ,
\eeq
as in \eq{Ci1I}.

Whereas these two correlation functions can be determined with just mean-current measurements, to determine $C_{32}$ we need to consider current cross-correlations.
Let us first imagine that we can measure the current through dot
$q$.  Then, in the limit $\Gamma_L \to 0$, such that there is only
ever at most one electron in the interferometer at a given
time, the zero-frequency noise correlator
\beq
   S_{3q' 2q} \equiv  \textstyle{\frac{1}{2}}\int dt\!\ew{\left\{I_{3q'},I_{2q}\right\}}_c
   ,
\eeq where $\ew{\ldots}_c$ denotes the cumulant average, is
proportional to the joint probability, $P_{3q' 2q}$, that the
electron travels through dot $q$ and ends up at output $3q'$. This
result follows in the same way as in Ref.~\onlinecite{Samuelsson2003};
the difference here being that we correlate the position of a
single electron in subsequent regions, as opposed to the
correlation of two spatially-separate electrons. Measuring all four such
correlators, we obtain the probabilities \beq
   P_{3q' 2q} = \frac{S_{3q' 2q}}{\sum_{r'r} S_{3r' 2r}}
   .
\eeq
From these directly-obtained probabilities, we construct the ideal-negative-measurement ones as
\beq
  P^\mathrm{INM}_{3q' 2q} = P_{3q'} - P_{3q' 2\overline{q}}
  ,
\eeq
where $\overline{q} =-q$ and the total probability at output $3q'$ is obtained from the currents
\beq
  P_{3q} = \frac{\ew{I_{3q}}}{\sum_r \ew{I_{3r}}}
  .
\eeq
These relations follow from charge conservation and the unidirectional nature of the transport.

The QPC detectors couple not the current flowing though the dot, but rather to their occupations.  In terms of the zero-frequency
correlation function between current fluctuations in the QPC and
those in one of the $3\pm$ ports, \beq
   S_{3q' Dq} \equiv \frac{1}{2} \int dt \! \ew{\left\{I_{3q'},I_{Dq}\right\}}_c
   ,
\eeq
the required probabilities read
\beq
  P_{3q' 2q} =
  \frac{\ew{\Delta I_{Dq}}S_{3q' Dq}}{\sum_{r'r}\ew{\Delta I_{Dr}} S_{3r' Dr}}
  .
\eeq This can be understood as follows.  Whereas $S_{3q' 2q}$
correlates two delta-function peaks corresponding to the passage
of the electron through the regions 2 and 3, $S_{3q' Dq}$
correlates a delta-function in region 3 with a signal of finite
duration in region 2, which corresponds to the finite time for
which the dot is occupied.  This mean occupation time is
proportional to the inverse of mean current through the dot, which
can be obtained (up to a proportionality constant) from the mean
detector current $\ew{\Delta I_{Dq}}$.

%%%
Calculating these probabilities, we find that in the limit
$\Gamma_L\to 0$, the third correlation function reads
\beq
  C_{32} &=& \cos \theta_B
  ,
\eeq
in accordance with \eq{C32angles}.  Since, in the
$\Gamma_L\to0$ limit, all three correlation function are identical
with their ideal counterparts, the LGI for this set-up is
identical to that of \eq{K3ideal}.
In the way that we have described the QPC detectors here, it does
not make any difference whether we calculate $K$ using the ideal
negative measurement probabilities or the direct ones since, in
our theoretical description, the QPC detectors act as ideal
detectors and only influence the system through their dephasing
effect.  Experimentally, the ideal negative measurement protocol
should be used, and actually, the comparison between the case with
ideal negative measurement and that without would give an
interesting method for studying to what extent the QPC measurements
are non-invasive.
Let us just add that, whilst the above results were derived in the
symmetric case with $\Gamma_{R+}/\Gamma_{R-} = 1$ and with balanced
detector rates, if these ratios are unequal but known, then the
difference can be accounted for by weighting the terms in the
correlation functions accordingly.

\subsection{Dephasing}

A simple way to include the effects of dephasing in this model is to ``leave the detectors switched on'' when calculating $C_{31}$. With empty and occupied rates
$\gamma_\mathrm{dephase}$ and $\gamma_\mathrm{dephase}'$, the measured $K$ function has the form of \eq{K3dephasing}, with the function $f(\Delta)$ replaced by
\beq
  f =
  \left[
    1 +
    \frac{
      \rb{
        (\gamma_\mathrm{dephase}')^{1/2} -  (\gamma_\mathrm{dephase}))^{1/2}
      }^2
    }
    {2\Gamma_R}
  \right]^{-1}
  .
\eeq $f=1$ is the ideal no-dephasing case, and $f=0$ gives the
classical limit.  To obtain strong violations of the LGI, therefore requires that the difference in rates $\gamma_\mathrm{dephase}'$ and $\gamma_\mathrm{dephase}$ is small compared with the tunnel rate $\Gamma_R$.

\subsection{Detection errors}

Just as the direct relation between the Bell inequalities and
noise measurements of, e.g.,
Refs.~\onlinecite{Samuelsson2003,Beenakker2003} relies on the weak-tunnel
limit\cite{Samuelsson2004,Lebedev2005,Beenakker2006,Samuelsson2009},
 so it is here that our measurements are only isomorphic with those required by the LGI in the $\Gamma_L\to0$ limit.
Away from this limit, there exists the possibility that our
measurements mistakenly correlate subsequent electrons, rather
than the same electron with itself.

\begin{figure}[tb]
  \begin{center}
\includegraphics[width=0.99\columnwidth,clip]{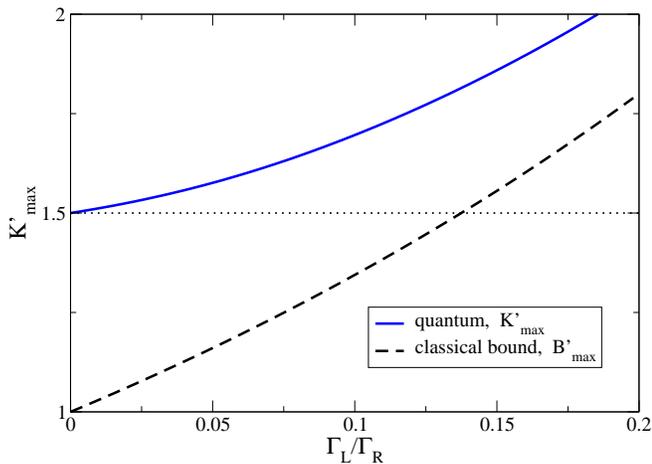}
  \caption{(Color online)
    The maximum value $K'_\mathrm{max}$ (blue solid line) and the corresponding classical value $B'_\mathrm{max}$ (black dashed line) away from the $\Gamma_L \to 0$ limit.  Both are higher than their ideal $\Gamma_L\to0$ values.
    A symmetric system was assumed and the fast detector limit $(\gamma'-\gamma)/\Gamma_R\to\infty$ taken.
    \label{DQD_GL_violations_fig}
 }
  \end{center}
\end{figure}

The LGI quantity can be calculated using the currents and
zero-frequency  noise, as described above, away from the
$\Gamma_L\to0$ limit to assess the error.
Assuming for simplicity that the detector is faster than the system dynamics  $\gamma'-\gamma\gg \Gamma_{L/R}$ (although the general case can easily be investigated too), we obtain for the LG correlator
\begin{widetext}
\beq
  K' &=&
    \frac{1}{(\Gamma_L-\Gamma_R) \Gamma_R}
    \left\{
        \left[-(\Gamma_L+\Gamma_R)^2+\Gamma_L (\Gamma_L+3 \Gamma_R) \cos^2\theta_A\right] \cos\theta_B
    \right.
    \nonumber\\
    &&
    \left.
        +(\Gamma_L-\Gamma_R) \Gamma_R \left[2 \cos\theta_A
        \sin^2(\theta_B/2)+\cos\phi \sin\theta_A \sin\theta_B\right]
    \right\}
    .
\eeq
\end{widetext}
This expression is again maximised with $\cos\phi = 1$, but, unlike
the $\Gamma_L\to0$ case, the maximizing angles $\theta_A$ and
$\theta_B$ are not equal.  If we assume that $\Gamma_L/\Gamma_R
\ll1$, we can expand to leading order ($\gamma'-\gamma\to\infty$)
to obtain
\beq
  K' &=& K +
  3\frac{\Gamma_L}{\Gamma_R}
  \cos \theta_B\sin^2\theta_A
  +O\!\rb{\rb{\textstyle{\frac{\Gamma_L}{\Gamma_R} }}^2}
  \nonumber
  ,
\eeq
where $K$ is the $\Gamma_L \to 0$ value.
We can also calculate the corresponding quantity in the
classical limit (this we do by calculating $C_{31}$ in limit $(\gamma'_\mathrm{dephase} - \gamma_\mathrm{dephase}) \to \infty$).   In this case, we obtain
$C^\mathrm{cl}_{31} = \cos \theta_A \cos \theta_B$, and the expansion of $B'=C_{21}+C_{32}-C^\mathrm{cl}_{31}$
for small $\Gamma_L$ gives 
\beq
  B' &=& B
  %\nonumber\\
  %&&
  +
    3\frac{\Gamma_L}{\Gamma_R} \cos \theta_B\sin^2\theta_A
   +O\rb{\rb{\textstyle{\frac{\Gamma_L}{\Gamma_R} }}^2}
  \label{K3classicalfiniteGL}
  ,
\eeq
where $B =  \cos \theta_A + \cos \theta_B - \cos \theta_A \cos \theta_B$ is the classical value in the ideal case which, when maximised gives $B_\mathrm{max}=1$, the bound of \eq{K3intro}.  Maximising $B'$ over the angles, we obtain a value bigger that unity.
To lowest order then, classical and quantum LG correlators
are affected in the same way.   \fig{DQD_GL_violations_fig} shows
the maximum values of both quantum and classical correlators.

Thus, assuming that we know the ratio of $\Gamma_L/\Gamma_R$ from
current and noise measurements, the effects of a finite tunneling
rate $\Gamma_L$ can be included in assessment of whether LGI is
violated or not.  The conservative approach is say that the
quantity $\left(K'_\mathrm{max} - \frac{3}{2}\right)$
represents a systematic  error in the measurement, and assuming
that this error works against us, we can only conclude that we
violate the LGI when the measured value of $K$ exceeds unity by
an amount equal to this error. Alternatively, one can say that
since one knows how the classical bound behaves at finite
$\Gamma/\Gamma_R$, we can simply use $B'_\mathrm{max}$ of
\eq{K3classicalfiniteGL} as a bound.  However, providing that we are in the
correct operating limit of $\Gamma_L / \Gamma_R \ll 1$,  these
modifications will be very small, such that whether they are taken into
account or not will only effect the question of violation in marginal cases.

%%%%%%%%%%%%%%%%%%%%%%%%%%%%%%%%%%%%%%%%%%%%%%%%%%%%%%%%%%%%%%%%%
%%%%%%%%%%%%%%%%%%%%%%%%%%%%%%%%%%%%%%%%%%%%%%%%%%%%%%%%%%%%%%%%%

%%%%%%%%%%%%%%%%%%%%%%%%%%%%%%%%%%%%%%%%%%%%%%%%%%%%%%%%%%%%%%%%%%%%%%%%%
%%%%%%%%%%%%%%%%%%%%%      CONCLUSIONS     %%%%%%%%%%%%%%%%%%%%%%%%%%%%%%%
%%%%%%%%%%%%%%%%%%%%%%%%%%%%%%%%%%%%%%%%%%%%%%%%%%%%%%%%%%%%%%%%%%%%%%%%%
\section{Conclusions \label{SEC:CONCS}}

We have considered the violation of the LGI in MZ inteferometer
geometries.  The key to the violation is a combination
of the interference at the
second beamsplitter and the inhibition of this interference by the
measurement process.
In the two proposals we have considered this inhibition occurs in two different ways. In the
first realisation, we physically interrupt transmission through
one of the arms of the MZI, obviously preventing interference.  On
the other hand, in the DQD proposal, the detectors act in a more
traditional way and introduce dephasing between the paths.

In this MZ geometry both the state of the electron and measurement time are 
mapped onto real-space coordinates --- the qubit states $Q=\pm$
are physically separate paths, and the regions within the
interferometer correspond to different time instances.  This
mapping has several advantages for seeking a violation of the LGI
in transport.  The mapping of the time-coordinate means that we do
not need to make time-resolved correlation measurements.
All the measurements required here
are either mean stationary currents or zero-frequency noise
correlators.  Furthermore, the spatial separation of the qubit
degrees of freedom facilitates the realisation of ideal negative
measurement, since it is relatively easy to couple to just one of
the qubit states when they are spatially distinct. In this
respect, increasing the separation of the detector arms should decrease
the plausibility of claims that detection in one arm is,
from a macro-realist point-of-view, influencing the other.

The general principles described here can easily be extended to
further systems.  Within transport, for example, our second scheme
could also be realised with  an edge-channel MZI plus QPC detector
channel without the quantum dots
\cite{Neder2007b,Neder2007a,Neder2007,Chang2008}.
An alternative setting for the realisation of our first scheme
might be the flying qubit experiment of Ref.~\onlinecite{Yamamoto2012}, which
is essentially a MZI away from the quantum Hall regime.  Two
challenges are obvious with this realisation. Firstly,  the leads
reported in the experiment have multiple channels, which
potentially gives rise to the problems discussed in section
\ref{SEC:MULTI}.  The second problem is that of
backscattering at the beamsplitters and detectors, which has
(justifiably) been neglected here but probably can not be
eliminated in set-ups such as that of
Ref.~\onlinecite{Yamamoto2012}.

Applications away from electronic transport are also possible. The
application of the first scheme in optics is obvious but the notion of the qubit state is predicated on the source being a single-photon source.  So whilst a classical
wave might also exceed the right-hand-side of \eq{K3intro}, this
would not constitute a violation of the LGI, as it represents an
application of the concepts outside their proper realm of
definition (i.e., non-dichotomic observables).
Going further, the same principles could be used to test the LGI with
electrons in free space, neutrons, atoms and molecules, all of
which have had interference experiments in the MZI geometry
conducted on them \cite{Schmiedmayer1997,Cronin2009}.  Of these,
molecules offer the most exciting prospect, as there the nature of
the coherence being tested could potentially be macroscopic, in
line with the original goals of Ref.~\onlinecite{Leggett1985}.

%%%%%%%%%%%%%%%%%%%%%%%%%%%%%%%%%%%%%%%%%%%%%%%%%%%%%%%%%%%%%%%%%%%%%%%%%

%%%%%%%%%%%%%%%%%%%%%%%%%%%%%%%%%%%%%%%%%%%%%%%%%%%%%%%%%%%%%%%%%%%%%%%%%
%%%%%%%%%%%%%%%%%%%%%%%%%%%%%%%%%%%%%%%%%%%%%%%%%%%%%%%%%%%%%%%%%%%%%%%%%
\begin{acknowledgments}
We are grateful to S.~Huelga, Y.~Ota, P.~Roche, P.~Samuelsson, and
M.~Yamamoto for useful discussions.  This work was supported by
the DFG through SFB 910. FN acknowledges partial support from the
ARO, JSPS-RFBR contract No. 09-02-92114, MEXT Kakenhi on Quantum
Cybernetics, and the JSPS-FIRST Program.
\end{acknowledgments}
%%%%%%%%%%%%%%%%%%%%%%%%%%%%%%%%%%%%%%%%%%%%%%%%%%%%%%%%%%%%%%%%%%%%%%%%%

%%%%%%%%%%%%%%%%%%%%%%%%%%%%%%%%%%%%%%%%%%%%%%%%%%%%%%%%%%%%%%%%%%%%%%%%%
%%%%%%%%%%%%%%%%%%%%%      REFERENCES     %%%%%%%%%%%%%%%%%%%%%%%%%%%%%%%
%%%%%%%%%%%%%%%%%%%%%%%%%%%%%%%%%%%%%%%%%%%%%%%%%%%%%%%%%%%%%%%%%%%%%%%%%
\bibliography{/home/gloygum/work/habilitation/bib/CEbib.bib,NLextra.bib}
%\bibliography{/home/cemary/work/habilitation/bib/CEbib.bib,NLextra.bib}

\end{document}